# CONCEPT OF QUATERNION – MASS FOR WAVE-PARTICLE DUALITY : A NOVEL APPROACH

R.C.Gupta
Professor, Institute of Engineering & Technology,
Lucknow, India
rcg_iet@hotmail.com

## ABSTRACT

In the present paper a new concept is introduced that: 'mass is a quaternion quantity'. The concept of quaternion mass suggests that the total mass M of a moving body is quaternion sum of : (i) the scalar (grain or rest) mass $m_g$ establishing its particle behavior and (ii) the vector mass $m_p$ governing its wave properties. Mathematically, the quaternion mass $M = m_g + m_p$ ; the magnitude $|M| = (m_g^2 + m_p^2)^{1/2}$ . The theory proposed here explains successfully several effects such as 'Compton effect' and 'refraction of light' which could not be explained otherwise by a single theory of wave or particle. Also explained are 'Doppler effect for light', 'photo-electric effect', 'Uncertainty principle' and 'Relativity'.

## 1. INTRODUCTION

The longest controversial issue in the scientific history is perhaps over the fundamental question of Nature that whether light is wave or particle? Huygens' wave theory succeeded over Newton's corpuscular theory as the wave theory explained successfully several important phenomena such as refraction and interference whereas particle theory could not. However, in the beginning of 20[th] century the particle theory of light (photon) emerged again as it explained clearly some observations such as photo-electric effect and Compton effect whereas the wave theory ailed to do so. In order to resolve the controversy, de-Broglie [1] proposed the hypothesis for wave-particle duality suggesting that light is both, particle and wave. Moreover he also suggested that a moving electron too can exhibit wave properties and it was experimentally proved to be so. However, the wave-particle dilemma especially for light still persists [2-6] in some ways. It is not cleat as to why electromagnetic radiation behaves as particle in one experiment and as wavy in another experiment. Both the particle and wave aspects have never been observed simultaneously, as if one aspect comes into being only if the other aspect is absent and vice-versa. Attempt to identify one aspect vanishes the other [2-6]. In an attempt to resolve the dilemma, the novel concept of 'quaternion mass' is presented and its applicability is demonstrated in this paper.

## 2. NEED FOR THE NEW CONCEPT OF MASS

Many arguments could be given in favor of the need for the new concept of mass. However, the author would state/quote a few, as follows:

(i) Mass and energy are no more separate quantities, but are interwoven by Einstein's mass-energy equation. This may imply that mass losses its independent identity and requires a new interpretation.

(ii) It is known, as Ugarov [7] mentions in his book on special relativity that 'rest mass of a system exceeds the sum of rest masses of constituent particles by a certain amount estimated in the reference frame in which the total momentum is zero'. He states further that 'rest mass is not additive quantity'. Such a property of mass is uncommon in classical mechanics. It is tempting to bring in some new definition of mass for constituent particle.

(iii) Is mass a scalar, a vector or a quaternion? Rest mass is definitely a scalar quantity. Photon, however has a zero rest-mass though it has a total mass $(h\nu/c)/c$ due to its momentum vector $h\nu/c$. What about the mass of a moving body having a rest-mass & momentum-vector? The quaternion was first proposed by Hamilton [8] as a sum of scalar plus vector [8-11]. Is mass 'a quaternion' for a moving body?

(iv) Both the aspects of a moving body i.e., the corpuscular (material) nature and its wave behavior have never been observed simultaneously. Any attempt to identify one aspect vanishes the other aspect, as if these two aspects come out of two different entities (say, scalar and vector). Could mass be considered as 'quaternion quantity', the scalar part of it representing the material-content (rest-mass), and the vector part (due to its momentum vector) governing its wave aspect?

(v) If introduction of a new concept of mass as 'quaternion quantity' could explain various phenomena and could eliminate/minimize ambiguity; it is worth doing. The new concept may appear to be speculative at first glance, but as it will be seen later that its success and capability to explain several diverse phenomena is striking. All avenues for the 'Truth' must be kept open.

## 3. QUATERNION-MASS CONCEPT

For the 'quaternion mass concept' introduced here it is proposed that 'mass is a quaternion quantity' and that the total mass M of a moving body has two components: (i) a scalar (particle at relatively-rest) 'grain-mass' $m_g$ and (ii) a vector 'photonic mass' $m_p$ due to its momentum. This may tentatively be considered as a postulate. However, it will be evident later that this novel concept of the 'quaternion-mass' has the potential to explain several phenomena without any real contradiction with the present status of formulation in physical sciences of concern.

The 'total quaternion-mass' M could be written as the quaternion sum of $m_g$ and $m_p$ as follows :

$M = m_g + m_p$                                                       (1.a)

When the particle-momentum is zero (particle at rest) its photonic (vector) mass is zero, whereas if its velocity is c (as for photon) its rest or grain (scalar) mass is zero. Taking one axis corresponding to the rest or gain mass and the other axis corresponding to the photonic mass, a diagrammatic representation of quaternion mass is suggested in Fig.1 .

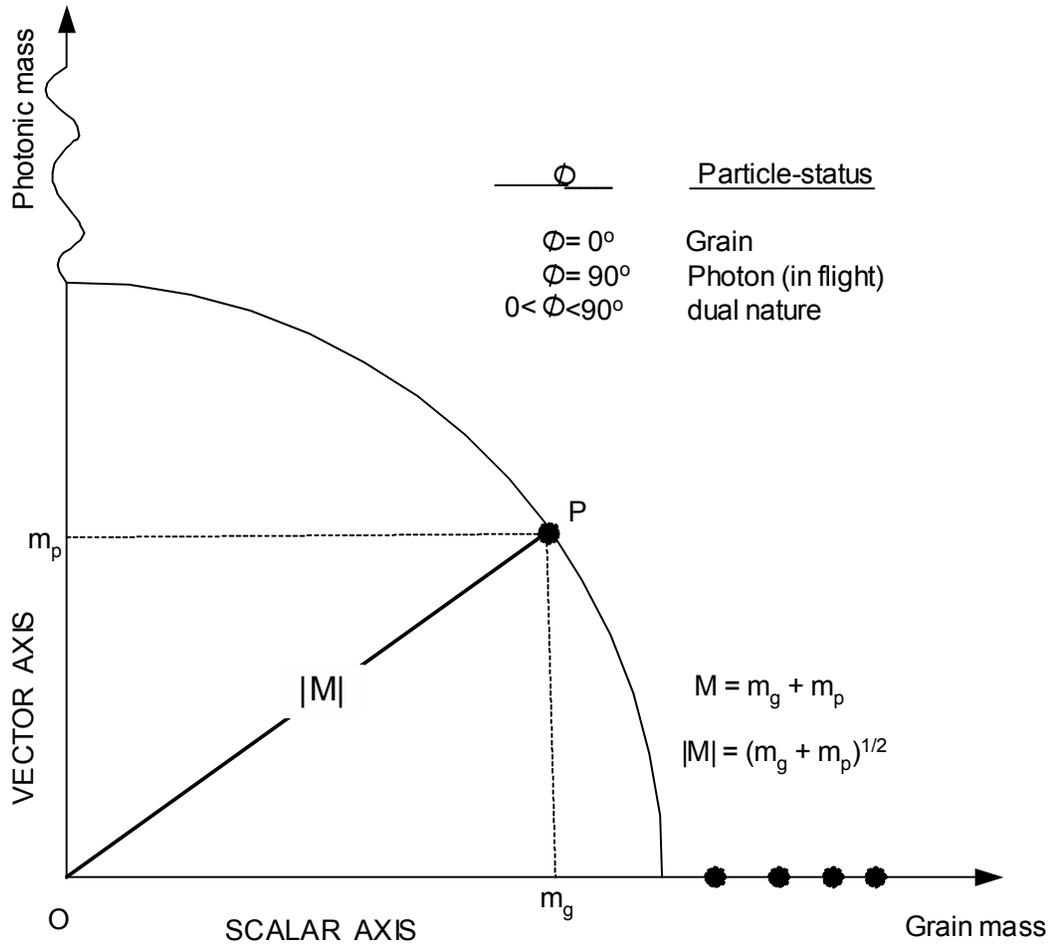

**Figure 1.**     **QUATERNION-MASS REPRESENTATION**

Total 'magnitude' |M| of the quaternion-mass M is 'Pythagorean-sum' of the constituents as follows as per property of the quaternion ,

$|M| = (m_g^2 + m_p^2)^{1/2}$                                               (1.b)

and the 'grain-photon phase angle' φ (Fig.1) is defined as,

$$\cos \varphi = m_g / |M| \quad \text{and} \quad \sin \varphi = m_p/|M| \qquad (1.c)$$

Particle's relative velocity is v and photon's speed is c. The 'total' momentum of the particle is $|M|.v$. It is considered that the rest (scalar) mass does not take any momentum and the total momentum is taken up by the photonic (vector) mass ($m_p$). Hence, $|M|.v = m_p.c$, or

$$m_p = |M|.v/c \qquad (1.d)$$

From equations (1.b) and (1.d) the following equation is obtained,

$$|M| = m_g / (1-v^2/c^2)^{1/2} \quad \text{or} \quad m_g = |M|.(1-v^2/c^2)^{1/2} \qquad (1.e)$$

The cross-relationship between the magnitude-of-mass $|M|$ and photonic-mass $m_p$ (as in Eqs. 1.b & 1.d) may appear to be interestingly intriguing showing dependence of one on the other, but is okay since it includes the relativistic effect as reflected in Eq.(1.e).

If $\varphi = 0°$, the particle is a scalar-particle (grain at rest) as in Fig.1; if $\varphi = 90°$, the particle is a vector-particle (photon in flight). For a particle moving with a velocity (v < c), $0° < \varphi < 90°$ and it has dual (quaternion) nature. Mathematically (from Eqs. 1.a, 1.d & 1.e),

$$M = m_g + m_p = m_g + |M|.v/c = |M| [\,(1-v^2/c^2)^{1/2} + v/c\,] \qquad (1.f)$$

**3.1 de-Broglie Wavelength**

Photonic-momentum p ($= m_p.c = |M|.v$) and photonic-energy E ($= h\nu$) are as usual related by $E = p.c = |M|.v.c$. Rest-mass does not take any momentum, total momentum is taken up by photonic mass $m_p$ leading to the de-Broglie wavelength λ (Eq.2) as under; as $p = |M|.v = m_p.c = E/c = h\nu/c$ and that $c = \nu.\lambda$, where h is the Planck's constant and ν is the frequency, λ is the wavelength & c is the speed of light,

$$\lambda = h / (\,|M|.v\,) \qquad (2)$$

**3.2 Mass-Energy Equation**

The total mass M of the particle when v = 0 is $|M|$, from Eq.(1.f), grain-mass as $|M|$ and photonic-mass as zero. If, however, the stationary mass converts itself into photonic-energy (wave) it may be considered as moving with a photonic-speed (v = c)

---

Concept of Quaternion-Mass....       Dr. R.C. Gupta                                  - 4 -

with grain or rest mass as zero and photonic mass as |M| leading to Einstein's mass-energy equation (Eq.3) as photonic energy $E = |M|.v.c$ ,

$$E = |M|.c^2 \qquad (3)$$

## 4. COLLISION OF PARTICLES AND CONSERVATION LAWS

Consider two particles 1 and 2, initial masses of which before-collision are $M_1 = m_{g1} + m_{p1}$ and $M_2 = m_{g2} + m_{p2}$. During collision, the grain and the photonic masses could be 're-distributed' between them. Consider that after-collision the new masses for the two particles are $M_1^/ = m_{g1}^/ + m_{p1}^/$ and $M_2^/ = m_{g2}^/ + m_{p2}^/$.

The total magnitude of the mass |M| for the colliding particles (which takes into account 'relativistically' both, the grain-mass and the photonic-mass) would therefore be 'conserved' giving the following equation,

$$|M_1| + |M_2| = |M_1^/| + |M_2^/| \qquad (4)$$

In the process of the collision of the particles, the total vector photonic mass is also considered to be 'conserved' and thus,

$$m_{p1} + m_{p2} = m_{p1}^/ + m_{p2}^/ \qquad (5)$$

In fact, as it will be more evident later in this paper that the conservation of total mass |M| (Eq.4) is the mass-energy conservation, and that the conservation of photonic mass $m_p$ (Eq.5) is the conservation of momentum.

It may be noted that the 'mass-quaternion' introduced in this paper takes into account the relativistic aspects as reflected in Eqs.(1.e & 1.f). Thus, in general, the rest (grain) mass is <u>not</u> conserved in collision unless the colliding particles retain their identities. It is also noted that the 'mass-quaternion' is <u>different</u> from the 'Hamiltonian's quaternion'; as mass-quaternion is in fact relativistic.

## 5. PHOTON SCATTERING

to study the photon scattering, take a general case of collision (Fig.2) where a photon strikes (at an angle x) on a moving object (say, electron) and scatters away. Consider that before-collision the electron moving with a velocity $v_1$ has a total mass $M_1 = m_g + |M_1|.v_1/c$ where $m_g$ is rest-mass of the electron, and that the incident photon (with a momentum $h\nu/c$) has a total mass $M_2 = 0 + h\nu/c^2$. After collision; the photon (at an angle θ, with momentum $h\nu^/$) is emerged with a total mass $M_2^/ = 0 + h\nu^//c^2$, and that the electron comes out (at an angle δ) moving with a velocity $v_1^/$ which has a total mass $M_1^/ = m_g + |M_1^/|.v_1^//c$.

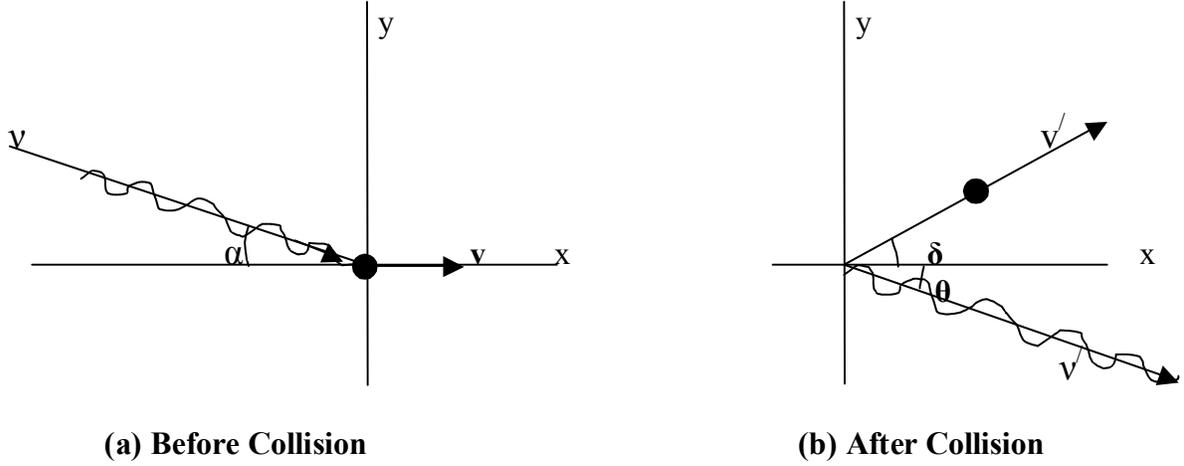

**(a) Before Collision**  **(b) After Collision**

**Figure 2. PHOTON SCATTERING**

Equation (5) for photonic-mass conservation reduces to the following equations of momentum conservation corresponding to the momentum conservation in x and y directions,

$$|M_1|.v_1 + h\nu/c \cos \alpha = |M_1'|.v_1' \cos \delta + h\nu'/c \cos \theta \qquad (6)$$

$$h\nu/c \sin \alpha = - |M_1'|.v_1' \sin \delta + h\nu'/c \sin \theta \qquad (7)$$

Equation (4) for total-mass conservation reduces to mass-energy conservation equation as,

$$|M_1|.c^2 + h\nu = |M_1'|.c^2 + h\nu' \qquad (8)$$

where from Eq.(1.e) magnitude of total of quaternion mass are as follows,

$$|M_1| = m_g/(1-v^2/c^2)^{1/2} \quad \text{and} \quad |M_1'| = m_g/(1-v_1'^2/c^2)^{1/2} \qquad (9)$$

From Eqs. (6),(7),(8) and (9), the following expression is obtained,

$$(\nu - \nu') = (h\nu.\nu' \cos \alpha)/(m_g.c^2) \cdot [1 - \cos(\theta-\alpha)] + v_1/c \cdot (\nu \cos \alpha - \nu' \cos \theta) \qquad (10)$$

For cases where the incident photon strikes not the electron but some heavier object such as an observer(or an observing instrument), mirror or glass; $m_g$ will be replaced with rest-mass of the object in Eq.(10) where the first term on R.H.S. would become negligible and the equation would reduce to,

$$(\nu - \nu') = (\nu \cos \alpha - \nu' \cos \theta) \cdot v_1/c \qquad (11)$$

### 5.1 Compton Effect

For $v_1 = 0$ and $\alpha = 0$, the Eq.(10) reduces to the equation for Compton effect [12] as follows,

$$(1/v' - 1/v) = h/(m_g \cdot c^2)\cdot(1 - \cos\theta) \tag{12}$$

### 5.2 Doppler Effect (For Light)

If the incident light (at $\alpha = 0$) reflects back (at $\theta = 180°$) from the moving object (mirror), the Eq.(11) reduces to,

$$v'/v = (1 - v_1/c) / (1 + v_1/c) \tag{13}$$

which gives the frequency of the reflected light($v'$) from an object – which is receding from the source (which emits light of frequency $v$). It should be noted that this frequency ($v'$) is Doppler-shifted twice, firstly when the moving object (mirror) receives the light and secondly when it reflects. The result obtained here is consistent with that predicted by Doppler-effect [13] for light.

### 5.3 Reflection and Refraction

### 5.3.1 Reflection

Consider that light of frequency $v$ strikes a stationary mirror ($v_1=0$) and that the frequency of the reflected light is $v'$. From Eq.(11), $v'=v$. Moreover, since free surface has zero shear-stress, momentum along free-surface should be same before & after impact. Mathematically, $hv/c \sin\alpha = hv'/c \sin\beta$ gives the reflection-law [14] as follows (incident angle $\alpha$ equals reflected angle $\beta$), since $v = v'$,

$$\alpha = \beta \tag{14}$$

### 5.3.2 Refraction

As in the case of reflection; for refraction of light into a stationary transparent medium, Eq.(11) gives $v'=v$. In case of reflection both the incident and reflected light travel in the same medium, but for refraction it is different. For refraction case, the incident light of frequency $v$ travels in a medium(say,air) where speed of light is $c$ and the refracted light of frequency $v'$ travels in another medium(say,glass) where speed of light is $c'$. Thus the momentum of the incident light is $hv/c$ whereas the momentum of refracted light is $hv'/c'$. Here again, since the interface between the two media can-not take shear stress, momentum conservation along the interface must hold good i.e., $(h\nu/c)\sin\alpha = (h\nu'/c')\sin\gamma$, which gives the Snell's law [14] for refraction as follows, since $\nu = \nu'$,

$$\sin\alpha / \sin\gamma = c/c' = \mu \qquad (15)$$

where $\mu$ is the refractive index of the second medium-material (say, glass $\mu > 1$) with respect to the first medium (say, air), which indicates that speed of light ($c'$) in denser-medium(glass) would be less than the speed of light (c) in rarer-medium(air), as expected and is in accordance with the results of wave theory.

It is to recall that the main reason for failure of Newtonian corpuscular-theory was that it predicts wrong results about speed of light in denser medium whereas the wave-theory predicts correctly. The quaternion-mass concept, which considers light as particle of photonic-mass (which differs in the two media, $h\nu/c^2$ in one medium and $h\nu'/c'^2$ in the other medium), also predicts (Eq.15) rightly.

### 5.3.3 Partial refraction and partial reflection

Consider that the incident light (n photons) strikes the stationary glass and that it is partially transmitted ($n_2$ photons) into and partially reflected ($n_1$ photons) back, such that $n = n_1 + n_2$. As shown earlier, the frequencies: of incident($\nu$), transmitted($\nu_t'$) and reflected($\nu_r'$) light are same. Here also, since the free-surface can-not take shear-stress, momentum conservation along the interface yields,

$$n(h\nu/c)\sin\alpha = n_1(h\nu_r'/c)\sin\beta + n_2(h\nu_t'/c')\sin\gamma$$

Taking $n = n_1 + n_2$ and $\nu = \nu_t' = \nu_r'$ the above equation reduces to

$$n_1[\sin\alpha - \sin\beta] + n_2[\sin\alpha - (c/c')\sin\gamma] = 0 \qquad (16)$$

For partial refraction and partial reflection, $n_2$ and $n_1$ are the positive integers thus Eq.(16) leads to both, the reflection-law (Eq.14) and the refraction-law (Eq.15). Possibly, specific combinations of $n_1$ and $n_2$ could lead to the understanding of bunching / anti-bunching of light [15].

### 6. PHOTO-ELECTRIC EFFECT

When a photon of energy ($h\nu$) strikes an electron(in an atom), part of its energy (W, the work-function) is used up for removing the electron from the atom, thus the remaining energy ($h\nu - W$) is used to give a velocity v to the electron.

If $m_e$ is the total mass of an electron in the atom, conservation of total mass (Eq.4) gives,

$$m_e + (h\nu - W)/c^2 = m_e/(1 - v^2/c^2)^{1/2} + 0$$

which gives the Einstein's equation for photo-electric effect [16] as follows (neglecting the higher order terms),

$$h\nu = \tfrac{1}{2} m_e v^2 + W \qquad (17)$$

Photon during its flight behaves as wave, but when it strikes an objects & it stops (either momentarily as in Compton-effect or completely as in photo-electric effect) the photon exhibits particle-like aspect.

## 7. UNCERTAINTY PRINCIPLE

Consider (in Fig.2) that a photon of frequency $\nu$ strikes (at $\alpha=0$) an electron and reflects back (at $\theta=\pi$) with a frequency $\nu'$ (from Eq.12 may be different from $\nu$ but as $h$ is very small, $\nu' = \nu$ ).

It can be shown from Eq.(6) that the magnitude of change (uncertainty) in the momentum (for $\alpha=0=\delta$, $\theta=\pi$) $\Delta mv = 2h\nu/c$.

Position of the electron from the photon-source S (Fig.2) is $x$. Distance traveled by the emitted photon to return back is $2x$. During this distance it may be considered that the light had N wavelength or $2x = N.\lambda$. The uncertainty in the measurement of position (or in N), would be of the order of an error of one wavelength, thus $\Delta x = \lambda/2$.

Therefore,

$$\Delta mv \cdot \Delta x = h \qquad (18)$$

which is of the same order of magnitude as that by the famous Heisenberg's uncertainty principle [17].

Here not only that it is shown that product of uncertainty in momentum and position is of the order of $h$ but also shown individually that error in position is half the wavelength and that error in momentum is twice the photon-momentum. For measurement with high frequency (small $\lambda$) photons, $\Delta x$ would be less but $\Delta mv$ would be more, product of these = $h$. A more rigorous consideration (scattering in all directions) would result this product as $h/(2\pi)$.

## 8. RELATIVITY

From Eqs. (1.b) and (1.d) the relativistic energy equation could be written as follows,

$$|M|^2 c^4 = m_g^2 c^4 + |M|^2 v^2 c^2 \quad (19.a)$$

$$E^2 = E_g^2 + p^2 c^2 \quad (19.b)$$

where total-energy $E = |M|.c^2$, relative-rest-energy $E_g = m_g.c^2$ and photonic-energy $E_p = |M|.v.c$ ; total momentum = photonic momentum = $m_p.c = |M|.v = p$.

Considering the momentum $|M|.v = |M|.(i\, v_x + j\, v_y + k\, v_z) = (i\, p_x + j\, p_y + k\, p_z) = p$, it could be shown from Eq.(19.b) that

$$(E_g/c)^2 = (E/c)^2 - (p_x^2 + p_y^2 + p_z^2) = (E'/c)^2 - (p_x'^2 + p_y'^2 + p_z'^2) \quad (20)$$

which is nothing but the well-known 4-vector invariant in unprimed and primed(/) coordinate system [18-20].

It may be noted that the relativistic energy equation (19) is derived from quaternion-mass concept equation (1). The concept of 'quaternion mass' finds strength from the fact that reverse is also possible i.e., the quaternion-mass concept equation (1) could be derived from the relativistic energy equation (19).

In fact, the 4-vector invariance (Eq.20) is equivalent to the invariance of rest (grain) mass, in unprimed and primed coordinate system. This is consistent with the quaternion concept of mass; because a change of coordinate system would affect velocity to change the vector photonic-mass thus changing the total-mass, whereas the scalar grain-mass (material) remains unchanged.

The special-relativity in the concept of quaternion-mass is also reflected in Eq.(1.e) and Eq.(3). in fact, the special-relativity is inherent in the concept of quaternion-mass.

## 9. DISCUSSION

To avoid ambiguity of wave-particle, some new terminology is suggested as follows. A particle at rest (in laboratory) is let called as 'grainon', a particle moving with a speed v (v < c) is let called as 'quaternion' and the particle moving with speed of light is called as 'photon'. All; grainon, quaternion and photon, let be 'said' as particles! However, the grainon (at relative-rest v = 0) has corpuscular (material) nature, the photon (in flight v = c) has the wave nature and the quaternion (in motion v < c) has both corpuscular & wave natures.

However, as per Eq.(3) inter-conversion of photon and grainon is possible. Moreover, when a photon is in motion (v = c) it behaves as wave (during its flight), but when it strikes an object it stops (v = 0, momentarily in 'Compton-effect' and completely in 'photo-electric effect' experiments) and thus the photon behaves as grainon. In experiments such as of 'interference' the two photon-streams interfere in the flight itself before coming to rest on screen, thereby showing the wave-nature, the essential-characteristic necessary for such experiments. That's why photon shows sometimes (in flight) wavy-aspect and sometimes (when it strikes another particle) corpuscular-nature.

Furthermore, an atom at rest may be apparently considered as grainon but it is composed of quaternion (electron) and grainon (nucleus), the nucleus in turn is composed of several particles of varied nature. In a way, one can say that an atom is a 'compound' particle.

The concept of quaternion-mass is new, interesting, promising and of fundamental importance. Its compatibility with special-relativity suggests that it is okay but at this stage there is no point in weighing this with full grown 4-vector special-relativity theory. However, a comparison could be made between the two for clarity as follows.

| **ITEMS** | **QUATERNION-MASS APPROACH** | **SPECIAL-RELATIVITY THEORY** | **REMARKS** |
|---|---|---|---|
| **Masses** | | | |
| Grain(rest) mass | $m_g$ | $m_o$ | $m_g = m_o$ |
| Photonic mass | $m_p = |M|.v/c$ | - | $m_p = (m^2 - m_o^2)^{1/2}$ |
| Total magnitude of mass | $|M|$ | $m$ | $|M| = m$ |
| **Momentum** | | | |
| Total momentum | $|M|.v$ | $m.v$ | |
| Photonic momentum | $m_p.c = |M|.v$ | - | |
| **Energy** | | | |
| Total energy | $|M|.c^2$ | $m.c^2$ | |
| Rest-mass energy | $m_g.c^2$ | $m_o.c^2$ | |
| Photonic energy | $|M|.v.c$ | - | |
| Kinetic energy | - | $mc^2 - m_o c^2$ | |
| **Conservation laws of** | | | |
| | $m_p$ | momentum | $m_p.c = m.v$ |
| | $|M|$ | mass energy | $|M|.c^2 = m.c^2$ |
| **Invariance of** | | | |
| | grain(rest) mass | 4-vector momentum | both are equivalent as in Eq.(20) |
| **Representation** | | | |
| Diagram | Quaternion representation(Fig.1) | Minkowskian space | |
| Dimensions | scalar(1) + vector(3) | space(3) + time(1) | both, 4-dimensions |

The author, however, like to stress that the quaternion-mass components are not merely a re-arrangement of relativistic masses but the important point is the introduction of the concept that mass of a moving body is a 'quaternion quantity' i.e., it has two components: (i) a scalar component as grain-mass (exhibiting the corpuscular nature) and (ii) a vector component as photonic-mass (governing the wave aspects). It may also, however, be noted that the 'mass quaternion' is not simply a conventional type of mathematical quaternion but is a new type of quaternion encompassing the relativistic aspects too, as reflected in Eq.(1.f).

Although several applications of quaternion-mass have already been considered earlier in depth, the author would cite the following example where the importance of photonic (vector) mass may possibly be further emphasized and usefulness of the quaternion-mass concept may be appreciated.

In quantum mechanical explanation of interference pattern of electrons in the famous double-slit experiment, we are forced to accept that 'an' electron passes through 'both' the slits simultaneously! Whereas in view of the 'quaternion' concept of mass; it may be considered that the electron (rest scalar mass) just-pass through one of slits only and the associated wave (vector photonic mass) split-passes through both the slits, or in other words the 'quaternion' divided into two parts, or the 'quaternion' divides into two different quaternion.

In the quaternion-mass concept, in fact, both the corpuscular and the wave aspects are merged together. The concept may provide a bridge-link between quantum-mechanics and special-relativity as well as between micro world and macro world. Special theory of relativity has close links to Electro-magnetism[21]. General theory of relativity [22] is the theory of Gravity, which tells that gravity is there because the 4-dimensional space-time is curved and that this curvature is because of presence of 'mass' there. Mass is in the central theme everywhere from atom to galaxy. What if the mass is a 'quaternion' quantity; the quaternion-mass is in accordance with special-relativity but how is it with general-relativity? Although at present stage it may appear that the concept of 'quaternion' mass raises more questions than it provides answers, but the concept is of fundamental importance and its full potentials will be realized in time to come.

## 10. CONCLUSIONS

The concept of 'quaternion' mass is new, simple and useful. It has the mathematical ingredients and the special-relativity is naturally embedded in it. It is able to explain phenomena which could not otherwise be explained by a single theory of wave or of particle. The wave-particle duality is mathematically incorporated in the quaternion-mass. The novel concept that 'mass is a quaternion quantity' is of fundamental importance and has the potential to explain several diverse phenomena.

ACKNOWLEDGEMENT

The author wishes to thank Dr.V.B. Johri, Professor-Emeritus, Lucknow University, Lucknow and Dr. M.S.Kalara, Professor, I.I.T. Kanpur for their useful discussions and comments. Thanks are also due to Shefali, Ruchi & Sanjay Gupta (Consultant CISCO, USA) for their assistance.